\documentclass{article}
\usepackage[utf8]{inputenc}
\usepackage[top=3cm,right=2cm,bottom=3cm,left=2cm]{geometry}
\usepackage{graphicx}
\usepackage{nomencl}
\usepackage{tikz}
\usetikzlibrary{matrix,shapes,arrows,positioning,chains,calc,automata,positioning,trees}
\usepackage{forest}
\usepackage{algorithm}
\usepackage{algorithmic}
\usepackage{mathtools,amssymb}
\usepackage{lastpage}
\usepackage{cleveref}
\usepackage{multirow}
\usepackage{booktabs,chemformula}

\crefmultiformat{equation}{(#2#1#3)}%
{ and~(#2#1#3)}{, (#2#1#3)}{ and~(#2#1#3)}
\crefname{equation}{equation}{equations}
\newcommand{\vect}[1]{\left( #1_1, #1_2, \,... \,, #1_n \right)}

\date{}
\title{Multi-Objective Optimization of Distribution Networks via Daily Reconfiguration}

\author{Seyed-Mohammad~Razavi, Hamid-Reza~Momeni, Mahmoud-Reza Haghifam, and~Sadegh~Bolouki}

\begin{document}

\maketitle

\begin{abstract}
			This paper presents a comprehensive approach to improve the daily performance of an active distribution network (ADN), which includes renewable resources and responsive load (RL), using distributed network reconfiguration (DNR). Optimization objectives in this work can be described as (i) reducing active losses, (ii) improving the voltage profile, (iii) improving the network reliability, and (iv) minimizing distribution network operation costs. The suggested approach takes into account the probability of renewable resource failure, given the information collected from their initial state at the beginning of every day, in solving the optimization problem. Furthermore, solar radiation variations are estimated based on past historical data and the impact of the performance of renewable sources such as photovoltaic (PV) is determined hourly based on the Markov model. Since the number of reconfiguration scenarios is very high, stochastic DNR (SDNR) based on the probability distance method is employed to shrink the scenarios set. At the final stage an improved crow search algorithm (ICSA) is introduced  to find the optimal scenario. The effectiveness of the suggested method is verified for the IEEE 33-bus radial distribution system as a case study.
		
		keywords: Distribution network reconfiguration, probability distance, solar generator, crow search algorithm.
	\end{abstract}
	
	\section*{Nomenclature}

	\begin{tabular}{p{2cm}p{14cm}}
 	$ \omega , \Omega  $  & Index and set of scenarios.\\
	$t , T$ & Index and set of time intervalss.\\
	$ g , G $ & Index and set of DGs.\\
	$ l, L $ & Index and set of DRs\\
	$ es, ESS $ & Index of ESS.\\
	$i , j , N_{bus}$ & Indices and total number of buses.\\
	$br , N_{br}$ & index and total number of branchs.\\
	$ re $ & Index of renewable resources. \\
	$ N_{s} $ & Total number of MCS.\\
	$ \rho^{Grid} $ & wholesale electricity price.\\
	$ SU , SD $ & Start-up and shut-down cost of DGs.\\
    $ \rho^{DR} $ & Contract price of DRs.\\
	$ L_{i,t} $ & Active power of load $ i $ in time $ t $. \\
	$ R_{br} $ & Resistance of branch $ br $. \\
	$ P_{g,max}$ & DGs maximum active power limits.\\ 
	$	P_{g,min}$ & DGs minimum active power limits.\\
	$ Z , F $ & Binary variables for DG commitment, start-up, and shut-down status. \\
	$ Q_{g,max}$ & DGs maximum reactive power limits.\\
	$ Q_{g,min}$ & DGs minimum reactive power limits.\\
	$ I_{br,max}$ & Branch maximum apparent capacity. \\
	$ \lambda , \mu $ & Failure rate and repair rate of renewable resources.\\
	$ P_{l,max} $ & Maximum active power contributed to DR.\\
	$ \rho^{SW} $ & price of switching.\\
	$ P_{re}^{D}$ & Probability of unavailable renewable resources.  \\
	\end{tabular}\\
\begin{tabular}{p{2cm}p{14cm}}
	$ P_{re}^{U}$ & Probability of available renewable resources. \\
	$ f(X)$ & Objective function. \\
	$ X $ & Control vector.\\
	$ B_{p,i} $ & Best position of crow $ i $.\\
	$ fl_{i} $ & Flight length of crow $ i $.\\
	$ AP_{i} $ & Awareness Probability of crow $ i $. \\
	$ IAP_{i} $ & Improved awareness Probability of crow $ i $. \\
\end{tabular}

\section{Introduction}
Today, world's energy systems are mostly supplied by conventional energy sources such as oil, gas, and coal, which together supply over 80\% of our primary energy. These fossil fuels contain a large amount of carbon, which is the leading cause of climate change  and global warming \cite{wang2014cost}. Climate change, persistent environmental pollution, and energy crises have made it more likely for many countries to use renewable energy instead of fossil fuels \cite{geels2015critical}. Challenges of managing traditional distribution systems are increasing with DGs such as PVs. Since the production of these resources, as opposed to conventional sources, is highly uncertain, distribution system management (DSM) must consider solutions in the presence of these uncertainties \cite{gomes2005new}.

Among notable ways that the DSM can deal with these challenges is to reconfigure the distribution network. Distribution network reconfiguration (DNR) is done by changing the cosmetic structure of distribution networks by the status of tie and sectionalizing switches. DNR methods can be divided to two categories, static and dynamic \cite{li2019coordinating}. Static reconfiguration methods are those that are often performed for on annual, seasonal, monthly, or weekly bases. Much of the existing literature on static reconfiguration assumes that DNR is performed using remotely controlled switches. However, this is not feasible in practice since installing and controlling these switches are very expensive. This, among other reasons, make dynamic reconfiguration schemes much better options than their static counterparts  \cite{li2019coordinating,lei2017identification}.
	 
	 Renewable resources may experience failures in different parts of their components \cite{chuangpishit2013topology}. Such failures are of great importance since renewable resources are responsible for a significant portion of the generated energy in the network. In this work, we propose a dynamic reconfiguration scheme which does not utilize remotely controlled switches, while considering the probability of failure of the renewable farms.
	  
	\subsection{Literature Review }
	The primary purpose of reconfiguration in the literature has been minimizing network losses \cite{pilo2011optimal,paterakis2015multi}. However, goals such as balancing load demand and improving reliability indicators have also been addressed \cite{jahani2019multi,andervazh2013adaptive}. Over the past two decades, numerous methods have been proposed to address the DNR problem, a survey of which can be found in \cite{mishra2017comprehensive}. The uncertainty of load and generation is one of the critical factors that must be taken into account for DNR \cite{abbaskhani2019distribution}. In \cite{azizivahed2019energy}, the uncertainty of renewable sources is modeled by 24-hour scenarios. There, for each hour, an optimal configuration according to generated and demand is presented. In \cite{fu2018toward}, allowing for the uncertainty of renewable generation, the problem of DNR is examined in three stages, the evaluation stage, the time division stage, and the optimization stage. Accounting for the daily load curves of consumers, \cite{lopez2004online} conducts DNR aiming to reduce active losses. In \cite{chen2006optimal}, the problem of DNR is investigated by factoring in the daily load demand to improve reliability indices. Considering the spatial and temporal capabilities of autonomous electric vehicles, and their demand for charging, DNR is performed in \cite{guo2019impacts}. There, a mixed-numeric programming model is  proposed, so that network reconfiguration is compatible with charging and discharging of electric vehicles. Authors in \cite{capitanescu2014assessing} investigate the DNR problem with the aim of increasing DG penetration capacity in thermal and voltage constraints and using direct values of consumer demand at different times. Risk-based reconfiguration is suggested in \cite{larimi2016risk}, considering load and generation uncertainty in the presence of reward/penalty schemes. In \cite{kavousi2013optimal}, the two-point estimation method is used to incorporate uncertainties into the analysis. Authors in \cite{ding2015feeder} propose a multi-period DNR model to take the dynamic load demand behaviors into account. The daily DNR problem is then solved using the genetic algorithm (GA) method to minimize the total network losses by considering the optimal DG output over the next 24 hours. In \cite{haghighat2015distribution}, a method for determining the minimum network losses with uncertain load and renewable generation is presented. In particular, a mixed-integer two-stage robust optimization formulation and a decomposition algorithm are proposed to address the problem.

	\tikzstyle{decision} = [diamond, draw, fill=blue!20, 
	text width=4.5em, text badly centered, node distance=3cm, inner sep=0pt]
	\tikzstyle{block} = [rectangle, draw, fill=blue!20, 
	text width=4em, text centered, rounded corners, minimum height=4em]
	\tikzstyle{blockkk} = [rectangle, draw, fill=blue!20, 
	text width=10em, text centered, rounded corners, minimum height=4em]
	\tikzstyle{blockkkk} = [rectangle, draw, fill=blue!20, 
	text width=15em, text centered, rounded corners, minimum height=4em]
	\tikzstyle{blockk} = [rectangle, draw, fill=blue!20, 
	text width=18em, text centered, rounded corners, minimum height=4em]
	\tikzstyle{line} = [draw, -latex']
	\tikzstyle{cloud} = [draw, ellipse,fill=red!20, node distance=3cm,
	minimum height=2em]
	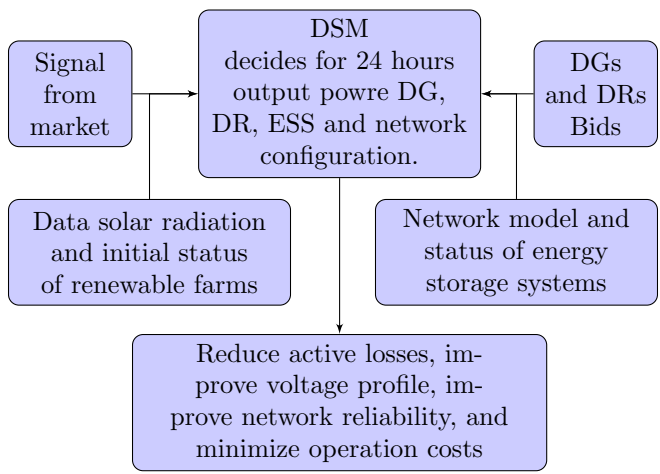
\begin{figure}
	\centering
		\begin{tikzpicture}[node distance = 2cm, auto]
		\node [blockkk] (init) {DSM \\decides for 24 hours output powre DG, DR, ESS and network configuration.};
		\node [block, left of=init , xshift=-45 ] (Signal from Market) {Signal from market};
		\node [blockkk, left of=init , xshift=-15, yshift=-60] (Data solar radiation) {Data solar radiation and initial status of renewable farms};
		\node [block, right of=init , xshift=40 ] (DGs and DRs Bids) {DGs and DRs Bids};
		\node [blockkk, right of=init , xshift=10 , yshift=-60] (Network model) {Network model and status of energy storage systems};
		\node [blockkkk, below of=init , yshift=-60] (o) {Reduce active losses, improve voltage profile, improve network reliability, and minimize operation costs};
		\path [line] (init) -- (o);
		\path [line] (Signal from Market) |- (init);
		\path [line] (Data solar radiation) |- (init);
		\path [line] (Network model) |- (init);
		\path [line] (DGs and DRs Bids) -- (init);
		\end{tikzpicture}
		\caption{Block diagram of the proposed work.}
		\label{fig.6}
	\end{figure}
	\subsection{Contributions}
	
	Major contributions of the paper are described below.
	
	$ 1) $ One of the most important issues for DSM is equipment failure since it can have a great impact on the network configuration.  In this work, as depicted in Figure. \ref{fig.6}, the effect of the probability of failure of renewable farms such as PV with uncertainty in solar radiation and the impact of DR and ESS on the result of the SDNR are explicitly investigated. Daily reconfiguration without remotely controlled switches is performed hourly depending on the load and generation. One of the main challenges when considering equipment failure is the very large number of scenarios. This leads to the need to develop random planning models. For this purpose, a stochastic distribution network reconfiguration is suggested.
	
	$ 2) $ In deriving static and dynamic reconfigurations, a variety of methods such as mathematical programming (e.g., mixed-integer linear programming \cite{borghetti2012mixed}, mixed-integer conic programming \cite{jabr2012minimum}, mixed-integer quadratically constrained programming \cite{romero2010simpler}, etc.) and metaheuristic techniques (e.g., GA \cite{ding2015feeder}, harmony search algorithm (HSA)\cite{rao2012power}, adaptive particle swarm optimization (APSO)\cite{malekpour2012multi}, etc.) have been used. The crow search algorithm is one of the newest exploratory algorithms. In this work, the improved crow search algorithm is used in the reconfiguration problem. 
	
	\subsection{Paper Organization }
	The rest of the paper is organized as follows. In section \ref{sec.2}, the objective function optimization is described. Section \ref{sec.3} explores the uncertainties of renewable resources. This section introduces a decision tree derived from resource failure and random behavior of solar radiation. The multi-objective crow search algorithm and proposed algorithm are presented in section \ref{sec.4}. Section \ref{sec.5}  presents a case study and the results obtained with the proposed analytical model. Finally, section \ref{sec.6} concludes the paper with discussions and future directions of this research.
	
	\section{Formulation of Multi Objective DNR } \label{sec.2}
	Reconfiguring the network is done with several decision variables. The decision variable $ X $ is a vector consisting of the tie and sectionalizing switches, the state of the DG, the demand response contract, and the energy storage system.
	\begin{align}
	\mathbf{X} & = \left[ \overline{T e i}, \overline{S W}, \overline{ D R}, \overline{E S S}\right]\nonumber\\
	\mathbf{\overline{T e i}} & = \vect{Tei}\nonumber\\
	\mathbf{\overline{S W}} & = \vect{SW}\nonumber\\
	\mathbf{\overline{D R}} & = \vect{DR}\nonumber\\
	\mathbf{\overline{E S S}} & = \vect{ESS}\nonumber.
	\end{align}
	\subsection{Objective Function }
	Based on the information received from the initial state of renewables at the time of optimization, the optimal objective function $ OF(X) $ is calculated as  
	\begin{align}
	&\textit{min}\:\mathbf{OF(X)} \!\!=\!\! \left[ OF_{1}(X),OF_{2}(X),OF_{3}(X),OF_{4}(X) \right]^\prime\nonumber,
	\end{align}
	where reliability, active losses, voltage deviation, and operating costs are the first to fourth parts of the objective function, respectively. these functions are briefly described below in that order.
	
	$ OF_1(X) $ represents the probability of hourly failure of the branches, which is an indicator for determining the reliability of the load point. Using the Monte Carlo simulation method, the probability of Failure distribution branches for hourly timing is modeled.
	The availability of branches per hour is denoted by failure time $ t_F $ and repair time $ t_R $, is calculated as \cite{conejo2010decision}
	\begin{align}
	t_F &\! =\! -MTTF\! \times\! \ln (u_1)\;\;,\;\;
	t_R \!=\!-MTTR\! \times\! \ln(u_2)\nonumber,
	\end{align}
	where $ MTTF $ and $ MTTR $ are the mean failure time and the mean
	repair time, respectively, and $ u_1 $ and $ u_2 $ are random variables that are uniformly distributed between 0 and 1. Values $ t_F $ and $ t_R $ are rounded to the nearest integer. Then, we have to update the time period ($ N_{T} $) and repeat this process until the the period is over \cite{conejo2010decision}. It is necessary to specify the failure time and the repair time by the binary parameters. For the branch $ br $ at time $ t $ with scenario $ s $, we define the variable $a_{br,t,s}$, to be equal to 1 in time $t_F$ and 0 in time $t_R$. 
	The probability of hourly failure of each branch is defined by the  $ 3D $ matrix $ R_{br} $,
	\begin{equation}
	\mathbf{R_{br}} = \begin{bmatrix}
	a_{1,1,1} &  \cdots & a_{1,1,N_s} \\
	\vdots  &  \ddots & \vdots  \\
	a_{N_{br},1,1} &  \cdots & a_{N_{br},1,N_s} \\
	\end{bmatrix}
	\nonumber.
	\end{equation}
	For each hour, a number between 0 and 1 is randomly generated. This number is multiplied by $ N_{T} $, and $ t $ is obtained. As a result, the reliability evaluated per hour is determined by the failure probability of active branches. Since the paper's focus is on the problem with network configuration, other equipment is definitely overlooked. The failure probability of any configuration is as follows:
	\begin{align}
	OF_1(X)=\sum_{i=1}^{N_{br}} L_{i,t}  \left( \frac{\sum_{s\in{S}}\,a_{br,t,s}\,x_{br}}{N_s} \right)\nonumber,
	\end{align}
	where $ x_{br} $ is a binary variable representing the active and inactive branch and  $ N_s $ is the number of scenarios implemented in the Monte Carlo simulation. We notice that Monte Carlo simulation is performed for all the branches before the optimization process begins and only the matrix $ A $ is used in the optimization. This means that the computational burden of the Monte Carlo simulation does not affect optimization.\\
	$ OF_2(X) $ represents active losses in the network and is expressed as
	\begin{align}
	OF_2(X)=\!\!\!\sum_{br\in{N_{br}}}\!R_{br} \left| I_{br} \right|^2\nonumber.
	\end{align}
	$ OF_3(X) $ represents of voltage deviation in each node, i.e.,
	\begin{align}
	OF_3(X) & =
	max\left[ \left| v_{ref} \! - \! min\:v_i \right|, \left| v_{ref}  \! -  \! max\:v_i \right|   \right]\nonumber.
	\end{align}
	Finally, $ OF_4(X) $ represents distribution network operation costs. It includes the cost of electricity purchased from the grid, the cost of energy generated by DG, the cost of switching, and the cost of using DRs, and is formulated as
	\begin{align}
	OF_4(X) & =C_{Grid}+C_{DG}+C_{DR}+C_{SW}\nonumber, 
	\end{align}
	where
	\begin{align}
	\label{eq.12}
	C_{Grid}=\rho^{Grid}P_{Grid},
	\end{align}
	\begin{align}
	\label{eq.13}
	C_{DG}\!=\!\sum_{g\in{G}}\!C_g+\sum_{g\in{G}} Z_{g} SU_g+\sum_{g\in{G}} F_{g} SD_g, 
	\end{align}
	\begin{align}
	\label{eq.14}
	C_{l}=\sum_{l\in{L}}\rho^{DR}P_{l}^{DR},
	\end{align}
	\begin{align}
	\label{eq.15}
	C_{SW}=\rho^{SW}\sum_{br\in{N_{br}}}\left|X_{br,t}-X_{br,t-1} \right|.
	\end{align}
	\Crefrange{eq.12}{eq.13} represent energy purchased from the grid and the cost of using DG, respectively \cite{jabbari2016microgrid,golshannavaz2014smart}. Here, $ C_g $ is calculated by $ a_gU_g+b_gP_g+c_gP_g^2 $, where $ a $, $ b $, and $ c $ are cost function coefficients. SU and SD are start-up and shut-down costs of DGs, respectively. Equation \eqref{eq.14} is the cost of power outages for consumers in DR \cite{golshannavaz2014smart}, and Equation \eqref{eq.15} is the cost of switching \cite{esmaeili2019optimal}. 
	\subsection{Electrical Constraints and Limits }
	The constraints used in the DNR problem are as follows:
	\begin{align}
	\label{eq.18}
	& V_{i.min} \leq V_{i} \leq V_{i.max}\;\;\;\;\; ,\;\;\;\;\;\;
	\;\;\;\;V_{slack}=1
	\end{align}
	\begin{align}
	\label{eq.19}
	& P_{g.min} \leq P_{g} \leq P_{g.max}\nonumber\\
	& Q_{g.min}\leq Q_{g} \leq Q_{g.max}
	\end{align}
	\begin{align}
	\label{eq.20}
	& SOC_{es}=SOC_{es,t-1}+\eta_{es}\left( P_{ES}^{-}-P_{ES}^{+} \right)\nonumber\\
	& SOC_{es.min} \leq SOC_{es} \leq SOC_{es.max}
	\end{align}
	\begin{align}
	\label{eq.21}
	& N_{br}=N_{bus}-1\nonumber\\
	& B=\left( I+M_G \right)^{N_{bus}-1}
	\end{align}
	\begin{align}
	\label{eq.22}
	& 0\leq P_{l} \leq P_{l,max}
	\end{align}
	\begin{align}
	\label{eq.23}
	& \left| I_{br} \right| <  I_{br,max}.
	\end{align}
	Equation \eqref{eq.18} represents the voltage constraints, while equation \eqref{eq.19} represents the active and reactive power limits DGs. Equation \eqref{eq.20} shows the charging and discharging limits of storages. Equation \eqref{eq.21} represents topology constraint. where $ M_G $ is the matrix adjacent to the configuration $ G $, and all Elements of $ B $ must be greater than zero $ (B>0) $. The major limitation in switching is that the structure in each configuration must remain radial. Equation \eqref{eq.22} represents DR constraints. According to the agreement between some consumers and DSM, energy consumption can be reduced to a limited extent. And finally, Equation \eqref{eq.23} represents the flow capacity of the branches. 
	
	\section{Scenario Generation and Reduction } \label{sec.3}
	Since DNR is an issue with a high computational complexity, other concurrent issues such as equipment failure are often overlooked. In this paper, we discuss possibility of failure of renewable resources as well as uncertainty in solar radiation simultaneously. Due the very large number of scenarios, we inevitably reduce them and create the so-called SDNR.
	\subsection{ Scenario Generation for Initial PV Status}
	In short-term studies, since the initial state of renewables is very influential, special attention must be paid to timing and initial state in order to calculate the probability of failure. Depending on the initial states for each renewable, the probability of different states during its 24 hours can be calculated using the Markov chain model. The probability of any renewable being available or unavailable at time t can be calculated as follows \cite{moshari2015short}:
	\begin{align}
	& P_{re}^{D}\!=\!\frac{\lambda_{re}}{\mu_{re}\!+\!\lambda_{re}}+\left( \mu_{re}\!\cdot\! P_{re}^{D}(0)\!-\!\lambda_{re}\!\cdot\! P_{re}^{U}(0) \right) \times Q \nonumber 
	\end{align}
	\begin{align}
	& P_{re}^{U}\!=\!\frac{\mu_{re}}{\mu_{re}+\lambda_{re}}\!+\!\left(\lambda_{re}\!\cdot\! P_{re}^{U}(0)\!-\!\mu_{re}\!\cdot\! P_{re}^{D}(0) \right) \times Q\nonumber
	\end{align} 
	\begin{align}
	Q = \frac{e^{-\left( \mu_{re}+\lambda_{re} \right) }}{\mu_{re}+\lambda_{re}}\nonumber. 
	\end{align}
	In this work, each PV generator consists of $ T $ number of renewables. Where only mode $ A $ is not available at $ t_0 $, the probability of any available mode $ B $ for each hour is calculated as \cite{moshari2015short}
	\begin{align}
	\pi_{av} &   = \!\!\!\sum_{\substack{i=0 \\ i\leq T\!-\!A \\ i \geq B-A}}^{B}\binom{T-A}{i}\binom{A}{B-i}P_{U0}^{{U_{t}^{i}}}\times P_{D0}^{{U_{t}^{B-i}}} \times P_{U0}^{{D_{t}^{T-A-i}}}\times P_{D0}^{{D_{t}^{A-B+i}}},\;\;\qquad\qquad av=0,...,T \nonumber.
	\end{align}
	The probability of the number of resources available per hour is  defined as
	\begin{equation}
	\begin{bmatrix}
	\pi_{1,0} &  \cdots & \pi_{1,T} \\
	\vdots  &  \ddots & \vdots  \\
	\pi_{24,0} &  \cdots & \pi_{24,T} \\
	\end{bmatrix}
	\nonumber.
	\end{equation}
	\subsection{ Scenario Generation for Solar Irradiance}
	In the literature, by collecting past historical data, solar radiation are modeled using beta probability density function (PDF) per hour of the day, which is as \cite{mazidi2014integrated,atwa2009optimal,hung2014determining}
	\begin{equation}
	f_{s}^{t}\left( S \right)  = 
	\begin{cases} 
	\frac{\Gamma \left( \alpha^{t}+\beta^{t} \right)}{\Gamma\left( \alpha^{t} \right)\Gamma\left( \beta^{t} \right) }S^{\left( \alpha^{t}-1 \right) } \left( 1-S \right)^{\left( \beta^{t}-1 \right) } , & 0 \leq S \leq 1\alpha,\beta\geq 0  \\ 0, &\mbox{otherwise}\nonumber.
	\end{cases}
	\end{equation}
	$ \alpha $ and $ \beta $ are distribution parameters, and are calculated as \cite{hung2014determining}
	\begin{align}
	\beta = \left( 1-\mu \right)\left( \frac{\mu \left( 1+ \mu \right)  }{\sigma^2}-1 \right)\;\;\;,\;\;\;\; \alpha=\frac{\mu \times \beta}{1-\mu}\nonumber, 
	\end{align}
	where $ \mu $ and $ \sigma $ are the mean and standard deviation from historical data, respectively. In the past, only one random variable was considered, and by comparing it to the cumulative distribution function (CDF), solar radiation was estimated. However, here we produce a larger set of random variables to increase the computational accuracy. The probability of each sample is $ \pi_{i}=\frac{n_{i}}{N} , i=1,...,R, $ where $ N $ and $ n_{i} $ are the total number of samples and the number of repetitions of sample $i$, respectively. Hence, the probability of solar radiation per hour is expressed as
	\begin{equation}
	\begin{bmatrix}
	\pi_{1,0} &  \cdots & \pi_{1,R} \\
	\vdots  &  \ddots & \vdots  \\
	\pi_{24,0} &  \cdots & \pi_{24,R} \\
	\end{bmatrix}
	\nonumber.
	\end{equation}
	\subsection{Scenario Reduction}
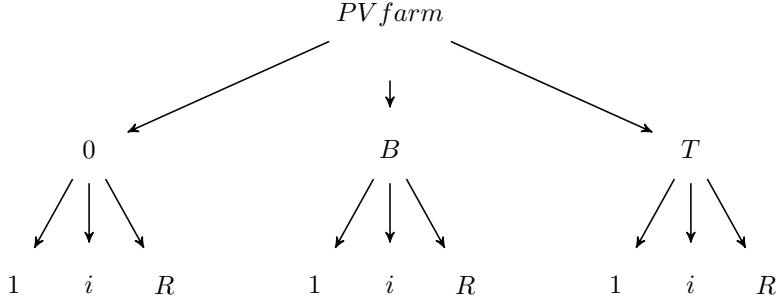
\begin{figure}
\begin{center}
	\begin{tikzpicture}[->,>=stealth',shorten >=3pt,auto,node distance=1.8cm,semithick]
	\tikzstyle{every state}=[fill=white,draw=none,text=black]
	\node[state] (A)  {$PV farm$};
	\node[state]         (B) [below of=A] {$B$};
	\node[state]         (D) [ right of=B,node distance=4cm] {$T$};
	\node[state]         (C) [left of=B,node distance=4cm] {$0$};
	\node[state]         (BB) [below of=B] {$i$};
	\node[state]         (BD) [ right of=BB,node distance=1cm] {$R$};
	\node[state]         (BC) [left of=BB,node distance=1cm] {$1$};
	\node[state]         (CB) [below of=C] {$i$};
	\node[state]         (CD) [ right of=CB,node distance=1cm] {$R$};
	\node[state]         (CC) [left of=CB,node distance=1cm] {$1$};
	\node[state]         (DB) [below of=D] {$i$};
	\node[state]         (DD) [ right of=DB,node distance=1cm] {$R$};
	\node[state]         (DC) [left of=DB,node distance=1cm] {$1$};
		\path (A) edge node{} (B);
		\path (A) edge node{} (C);
		\path (A) edge node{} (D);
		\path (B) edge node{} (BB);
		\path (B) edge node{} (BC);
		\path (B) edge node{} (BD);
		\path (C) edge node{} (CB);
		\path (C) edge node{} (CD);
		\path (C) edge node{} (CC);
		\path (D) edge node{} (DB);
		\path (D) edge node{} (DC);
		\path (D) edge node{} (DD);
	\end{tikzpicture}
\end{center}
		\caption{Decision tree diagram for a solar farm}
		\label{fig.8}
\end{figure}

	According to the scenarios related to the initial state and solar radiation using stochastic programming, a decision tree like Figure. \ref{fig.8} can be formed with three steps and  $ N_{\varOmega} = T \times R $ scenarios.
	We notice that it could be difficult to use this method to solve a daily optimization problem since it is not scalable, meaning that increasing the number of scenarios will significantly increase computation time. Therefore, we employ a probabilistic distance method to reduce the number of scenarios. The most common probability distance utilized in stochastic programming is the Kantorovich distance \cite{conejo2010decision}. First, function $ \nu $ is defined as the norm of the difference between pairs of scenarios, that is \cite{conejo2010decision}
	\begin{align}
	\nu\left( \omega , \omega^\prime \right) & = \left|\!| r\left( \omega \right) -  r\left( \omega^\prime \right)  |\!\right|, \qquad ~~ \forall \omega\in{\varOmega}\nonumber,   
	\end{align}
	where $ r(.) $ is the outcome of each scenario at each step of the decision tree. The values of function $ \nu $ can be conveniently arranged into a symmetric matrix with zero diagonal elements, where each row (and column) represents a scenario. We now perform an iterative algorithm starting with the set of all possible scenarios $ \varOmega_{j}=\left\lbrace  1,2,\ldots , N_{\varOmega} \right\rbrace $. We find the scenario $ \left(  \omega_{s}\right)  $  within $\Omega_j$, which has the minimum aggregated distance to other scenarios in $\Omega_j$ \cite{conejo2010decision}, that is
		\begin{align}
d_{\omega} & = \sum_{\omega^\prime = 1 , \omega^\prime \ne \omega}^{N_{\varOmega}} \pi_{\omega^\prime} \nu\left( \omega , \omega^\prime \right), \qquad\qquad\; ~~ \forall \omega\in{\varOmega_{j}},\nonumber\\
\omega_{s} & = \in{\arg_{\omega \in \varOmega_{j}}{\textit{min}\;{d_{\omega}} }}\nonumber.
	\end{align}
We then update values of the matrix and remove $\omega_s$ from the set $\Omega_j$ of scenarios, i.e.,
\begin{align}
\nu\left( \omega , \omega^\prime \right)  = & \min \left\lbrace \nu \left( \omega , \omega^\prime \right) , \nu \left( \omega , \omega_{s} \right) \right\rbrace,  ~~ \forall \omega , \omega^\prime \in{\varOmega_{j}},\nonumber\\
\varOmega_{j}  = & \varOmega_{j} \backslash \omega_{s}\nonumber.
\end{align}	
After a number of iterations, the set $ \varOmega_{s}^{*} = \varOmega \backslash \varOmega_{j} $ is selected as the set of preferred scenarios. Redistribution of probabilities can be accomplished as follows. The probabilities of selected scenarios $ \omega \in{\varOmega_{s}^{*}} $ are computed as $ \pi_{\omega}^{*} \longleftarrow  \pi_{\omega} + \sum_{\omega^\prime \in{J(\omega)}}\pi_{\omega^\prime}$, where $ J(\omega) $ is defined as the set of scenarios $ \omega^\prime \in{\varOmega_{j}^{\left[ 2 \right]}} $ so that $ \arg{\textit{min}_{\omega^{\prime\prime}\in{\varOmega_{s}}} \nu \left( \omega^{\prime\prime} , \omega^{\prime} \right)}$. The pseudo-code probability distance is below \cite{conejo2010decision}.

\begin{algorithm}
		\caption{probability distance}
		\begin{algorithmic}
			\STATE Compute function $ \nu^{\left[ 0 \right] } \left( \omega , \omega^\prime \right) $ for each pair of scenarios $ \omega $ and $ \omega^\prime $ in $ \varOmega $.
			\STATE Set $ \varOmega_{j}^{\left[ 0 \right] }= \left\lbrace  1,2,\ldots , N_{\varOmega} \right\rbrace$
			\FOR { $ i = 1 $ to $i_{Max}$ }
			\STATE Compute $ d^{i}_{\omega} $ 
			\STATE Select $  \omega_{ i } \in{\arg_{\omega \in \varOmega}{\textit{min}\;{d_{\omega}^{\left[ i \right]}} }} $
			\STATE Set $ \varOmega_{j}^{\left[ i \right] }= \varOmega_{j}^{\left[ i-1 \right] }\backslash \omega^{\left[ i \right]}$
		    \STATE	updated $ \nu^{\left[ i \right]}\left( \omega , \omega^\prime \right) $
			\ENDFOR	
			\STATE Compute $ \pi_{\omega}^{*} $
		\end{algorithmic}
\end{algorithm}
	Since we have considered both failure and solar radiation, a scenario is represented by a pair $(av^{ch}),(i^{ch})$, where $av^{ch}$ indicates the failure scenario and $i^{ch}$ indicates the solar radiation scenario. Given $F$ renewable resources, the selection probabilities of scenario $(av^{ch}),i^{ch})$ in 24 hours of the day can be represented as the following matrix:
	\begin{equation}
	\begin{bmatrix}
	\pi_{av^{ch},i^{ch}}^{1,1} & \cdots &  \pi_{av^{ch},i^{ch}}^{1,F} \\
	\vdots  & \ddots & \vdots  \\
	\pi_{av^{ch},i^{ch}}^{24,1} & \cdots &  \pi_{av^{ch},i^{ch}}^{24,F} \\
	\end{bmatrix}
	\nonumber.
	\end{equation}
	\subsection{Solar Generation} \label{sub.1}
	The output power of the solar generator depends on the temperature and the radiation of the sun which are calculated as \cite{hung2014determining}
	\begin{align}
	P_{pv_{s}} & = \gamma \times FF \times V_{ij} \times I_{ij}\nonumber\\
	T_{ct} & = T_{at} + s \times \left\lbrace \frac{NOCT-20}{0.8} \right\rbrace \nonumber\\
	V_{ij} & = V_{oc} - K_{vt} \times T_{ct}\nonumber\\
	I_{ij} & = s \times \left\lbrace I_{sc} + K_{ct} \times \left( T_{ct}-25 \right)  \right\rbrace \nonumber\\
	FF & = \frac{V_{mp} \times I_{mp}}{V_{oc} \times I_{oc}}\nonumber,
	\end{align}
	where $ P_{pv_{s}} $ is output power at solar irradiance $ s $, $ \gamma $ is the number of solar modules, $ V_{ij} $ and $ I_{ij} $  are the output voltage and current, $ T_{ct} $  and $ T_{at} $ are the module temperature and the ambient temperature at which the module is located  $ \left(  ^{\circ}\mathrm{C} \right)  $, $ NOCT $ cell temperature in nominal operation, $ s $ is solar radiation rate, $ K_{ct} $ and $ K_{vt} $ is the voltage $\left(  V/^{\circ}\mathrm{C} \right)  $ and current $ \left(  A/^{\circ}\mathrm{C} \right) $ temperature coefficients, $ V_{mp} $ and $ I_{mp} $ are the maximum voltage and current power point, and finally $ V_{oc} $ and $ I_{oc} $  are the open-circuit voltage and short circuit current.
	\section{Multi Objective Optimization }\label{sec.4} 
	In general, a multi-objective optimization problem with different constraints can be expressed as
	\begin{align}
	\textit{min}\:\mathbf{OF(X)} & = \left[ F_{1}(X) \, , \, F_{2}(X) \,  , \, ... \, , \, F_{n}(X)  \right]^\prime\nonumber\\
	\mathbf{h_{i}(X)} & = 0 \, , i = 1 \,,2\,,...\,,N_{eq}\nonumber\\
	\mathbf{g_{i}(X)} & \leq 0 \, , i = 1 \,,2\,,...\,,N_{ueq}\nonumber,
	\end{align}
	where $ \mathbf{h_{i}} $ and $ \mathbf{g_{i}} $ are the equality and inequality constraints, respectively. In multi-objective optimization, the objective functions may conflict, leading to inevitable tradeoffs. Hence, the search for non-dominated solutions called Pareto optimal solutions is considered. For a Pareto optimal solution, at any point in the space of the problem, no point can be found to improve each and all objective functions. In other words, solution $ \mathbf{X} $ dominates solution $ \mathbf{Y} $, if $ \mathbf{Y} $ is by no means better than $ \mathbf{X} $, while $ \mathbf{X} $ is in at least one case better than $ \mathbf{Y} $, i.e:
	\begin{align}
	\forall{j} & \in 1\,,2\,,...\,,N_{f}\,,F_{j}\mathbf{(X)}\leq F_{j}\mathbf{(Y)}\nonumber\\
	\exists{k} & \in 1\,,2\,,...\,,N_{f}\,,F_{k}\mathbf{(X)}< F_{k}\mathbf{(Y)}\nonumber.
	\end{align}
	Since the DNR problem is a nonlinear optimization problem with equality and inequality constraints, it needs to be converted into an unconstrained one by constructing an augmented objective function incorporating penalty factors for any value violating the constraints:
	\begin{align}
	F_{i}\left( X \right) & =f_{i}\left( X \right) +  p_{1} \sum_{j = 1}^{N_{eq}} \left( h_{j}\left( X \right) \right)^2 +  p_{2}  \sum_{j = 1}^{N_{ueq}}  \left( Max \left[ 0\,,g_{j} \left( X \right) \right]  \right)^2\nonumber.
	\end{align}
	Given that the aforementioned functions have different properties, we use fuzzy membership to formulate them on an equal footing, that is
	\begin{equation}
	\label{eq.50}
	\mu f_{i} = 
	\begin{cases} 
	1, & f_{i}\left( X \right) \leq f_{i}^{min} \\  0, & f_{i}\left( X \right) \geq f_{i}^{max} \\ \frac{f_{i}^{max} - f_{i}\left( X \right)}{f_{i}^{max} - f_{i}^{min}}, & f_{i}^{min} \leq f_{i}\left( X \right) \leq f_{i}^{max}\nonumber,
	\end{cases} 
	\end{equation}
	where $ f_{i}^{max} $ and $ f_{i}^{min} $ are the upper and lower limits of each of the objective functions, respectively, calculated according to system constraints.
	\section{Solution Procedure } \label{sec.5}
	In this section, the proposed Improved Crow Search Algorithm (ICSA) is described in details.
	\subsection{Original Crow Search Algorithm (CSA) }
	The CSA is one of the newest optimization methods introduced by Askarzadeh \cite{askarzadeh2016novel}. This method is inspired by the crows' intelligent behavior in hiding their food to solve the optimization problem. This algorithm provides a simple concept and effective technique that can	be implemented quickly. In this algorithm, the crow $ i $ flight length in each iteration is shown by $ (fl) $ as well as the degree of awareness of the crow $ j $ pursuit by awareness probability $ (AP) $. The pseudo-code CSA is below.
	\begin{algorithm}
		\caption{The original CSA}
		\begin{algorithmic}
			\STATE Initialize problem crow and adjustable parameters
			\STATE Evaluate the objective function and determine best position $ (B_{p}) $ for each crow
			\WHILE {$ It = It_{Max} $ }
			\FOR {$ i = 1 \; \textit{to nPop} $}
			\STATE The crow $ i $ randomly selects crow $ j $ from the population.
			\IF {$ rand_{i} \geq AP_{j} $}
			\STATE $ \mathbf{X}_{i}^{it} = \mathbf{X}_{i}^{it-1} + rand_{i} \times fl_{i} \times \left( B_{p,j}^{it-1} - \mathbf{X}_{i}^{it-1} \right)   $
			\ELSE
			\STATE $ \mathbf{X}_{i}^{it} =  \textit{a random position of search sapce} $
			\ENDIF
			\ENDFOR	
			\STATE Check the feasibility of new positions
			\STATE Evaluate the objective function
			\STATE Update the best position of each crow 			
			\ENDWHILE
		\end{algorithmic}
	\end{algorithm}
	\subsection{Improved Crow Search Algorithm (ICSA)}
	\subsubsection{Improved Awareness Probability}
	In the standard CSA, by reducing the $ AP $ value, the algorithm performs a search in a local area. As a result, this leads to finding a right solution in that area. On the other hand, if the AP value increases, the algorithm performs a global search. The awareness probability parameter in the standard CSA is a fixed number. However, if the $ AP $ remains constant in the optimization process, it may not produce the desired results. In fact, by changing the parameter of the $ AP $ to the improved awareness probability $ (IAP) $, the ratio between diversification and intensification can be controlled. In this paper, crow $ j $ and crow $ i $ use external memory and the $ AP $ parameter is set according to this memory. In this setting, if the crow $ j $ has a better response to the objective function (better memory), the crow $ i $ should have a better chance of chasing it. As a result, the $ IAP $ can be seen as follows:
	\begin{align}
	IAP_{j,it} &\!\! =\!\! u_{ap} \!\! \times \!\! \left(\!\! \frac{max\! \left( \mu f_{j,k}^{it}\right)}{min\! \left( \mu f_{i,k}\right) } \!\!\right) \!\! \times \!\! \alpha_{ap} \;,\; k=1,...,4 \nonumber,
	\end{align}
	where $ u_{ap} $ is a random number between 0 and 1, and $ \alpha_{ap} $ is a fixed number indicating the improved awareness probability.\\
	\subsubsection{Levy Distribution}
	Levy distribution is used in this work for the random search of each crow. Levy Flight is a powerful mathematical tool introduced by Paul Levy. In the search space, Levi's distribution is usually more efficient than a uniform random distribution \cite{jain2017improved}. First, compute $ Z_{i} $ as a step size using the Mantegna method \cite{yang2010nature}:
	\begin{align}
	Z_{i} & = \frac{r_{a}}{|r_{b}|^{\frac{1}{\tau}}} \;\;\;\; , \;\;\; \;
	r_{a}  \sim N \left( 0 , \sigma_{a}^{2} \right) \;\;,\;\; r_{b}  \sim N \left( 0 , \sigma_{b}^{2} \right) \nonumber\\
	\sigma_{a} & = \left( \frac{\Gamma \left( 1 + \tau \right) \times \sin{ \left( \pi \tau\backslash 2 \right)  } }{\Gamma \left( 1 + \tau \backslash 2 \right) \times \tau \times 2^{\tau-{1 \backslash 2}}  } \right)^{1 \backslash \tau } \;\;,\;\; \sigma_{b} = 1\nonumber.
	\end{align}
	Now with $ Z_{i} $ we can calculate the new crow position as
	\begin{align}
	\mathbf{X}_{i}^{it} & \!=\! \mathbf{X}_{i}^{it-1}\! +\! fl \! \times \! \left(\mathbf{B}_{p,i} \!-\! \mathbf{X}_{i}^{it-1}  \right)\! + \!Z_{i}\! \times\! \left(\mathbf{X}_{l}^{it}\! -\! \mathbf{X}_{i}^{it-1}  \right)\nonumber,
	\end{align}
	where $ \mathbf{X}_{l}^{iter} $ is one of the Pareto solutions in the repository and is calculated using the Roulette Wheel. We note that to avoid exceeding computational burden and memory limited, the size of the repository is defined to be constant.
	\section{Numerical Simulations} \label{sec.6}
	In this section, the performance of the proposed method is examined in the 33-bus IEEE test system. \cite{baran1989network} provides information on the network 33-bus. The constraint of operating voltages is assumed from 0.95 to 1.05 p.u. Here, according to \cite{lopez2004online}, we consider different types of consumers such as industrial, commercial and residential with different demand time. Table \ref{tab.10} shows the connection points of each. Also, in \cite{ghofrani2014distribution}, the energy price is given in 24 hours. The RLs are at nodes 8, 14, 25, 26, 31 and 33. Under the agreement, the DMS can control up to 50\% of consumption.
	\begin{table}	
	\centering
		\caption{Cost coefficients and technical data for DG units}
		\label{tab.1}
		\begin{tabular}{ *{8}{c} }
			\toprule
			& $ a_{g}  $ &  $ b_{g} $ & $ c_{g} $ & $ Su_{g}  $ & 	$ Sd_{g} $ & $ P_{g,max}\left( KW \right)  $ & $ P_{g,min}\left( KW \right)  $ \\
			\midrule
			\textbf{DG1} & 27 & 79 & 0.0035 & 15 & 10 & 1000 & 100 \\
			\textbf{DG2} & 25 & 87 & 0.0045 & 15 & 10 & 800 & 80 \\   
			\bottomrule
		\end{tabular}
	\end{table}
	\subsection{Assumptions} 
	We assume that there are five renewable farms (solar generator) at nodes 10, 21, 25, 30, and 32. Each of them has five members with the same features. Historical data (mean and standard deviation) of solar radiation for five 1MW PVs are given in \cite{mazidi2014integrated}. Their features are listed in Table \ref{tab.3}. Two DGs with capacities of 1MW and 0.8MW are located at nodes 15 and 18. It is also located at nodes 10 and 30 of the ESS. Their features are listed in Tables \ref{tab.1} and \ref{tab.4} \cite{golshannavaz2014smart}.
	
	The distribution system includes five branches of maneuver tie-lines. All branches are equipped with controllable switches, and each switch can participate in switching operations for reconfiguration up to 4 operations a day.
	\begin{table}	
	\centering
		\caption{The PV characteristic}
		\label{tab.3}
		\begin{tabular}{ *{6}{c} }
			\toprule
			\multicolumn{6} {c} {\textbf{PV}} \\
			\midrule
			$ T_{at} $	 & $ 30 \left(  ^{\circ}\mathrm{C} \right) $ & 	$ V{mp} $ & $ 18  \left( V \right) $ & 	$ K_{vt} $	 & $ 0.38  \left( \% A \right)  $  \\
			$ I_{mp} $	 & $ 11.12  \left( A \right)  $	    & $ \gamma $     & $ 1000  $ & 	$ P_{rate} $ & $ 200  \left( W \right)  $  \\
			$ V_{oc} $	 & $ 22.30  \left( V \right)  $	    & $ M $	         & $ 5 $     & $ K_{et} $	 & $ 0.1  \left( \% A \right)  $  \\
			$ I_{sc} $	 & $ 11.89  \left( A \right)  $	    & failure rate & $ 144 $     & repair rate	 & $ 5 $  \\				
			\bottomrule		
		\end{tabular}
	\end{table}
	The reliability information of distribution branches is given in Table \ref{tab.5}.
	Also, the costs of PVs and ESS are assumed to be negligible.
	\begin{table}
	\centering
		\caption{The ESS unit characteristic}
		\label{tab.4}
		\begin{tabular}{ *{4}{c} }
			\toprule
			$ 	SOC_{max}\left( KW \right) $ & 	$ 	SOC_{min}\left( KW \right) $ & $ P_{max}^{+}\left( KW \right) $ & $  P_{max}^{+}\left( KW \right) $   \\
			\midrule
			300	 &  80	   & 40	 &  40\\
			\bottomrule
		\end{tabular}
	\end{table}
	\begin{table}
	\centering
		\caption{demand patterns of each load point}
		\label{tab.10}
		\begin{tabular}{ *{13}{c} }
			\toprule
			\textbf{Demand pattern} & \multicolumn{12}{c}{Node} \\
			\cline{2-12}
			\toprule
			residential & 12 & 13 & 14 & 15 & 16 & 17 & 28 & 29 & 30 & 31 & 32 & 33 \\
			commercial & 6  & 7  & 8  & 9  & 10 & 11 & 25 & 26 & 27 & -  & -  & - \\
			industrial & 2  & 3  & 4  & 5  & 18 & 19 & 20 & 21 & 22 & 23 & 24 & - \\
			\bottomrule
		\end{tabular}
	\end{table}
	\subsection{Comparison of ICSA With Other Methods} 
\begin{figure}
		\begin{center}
			\includegraphics[height=6cm,width=10cm]{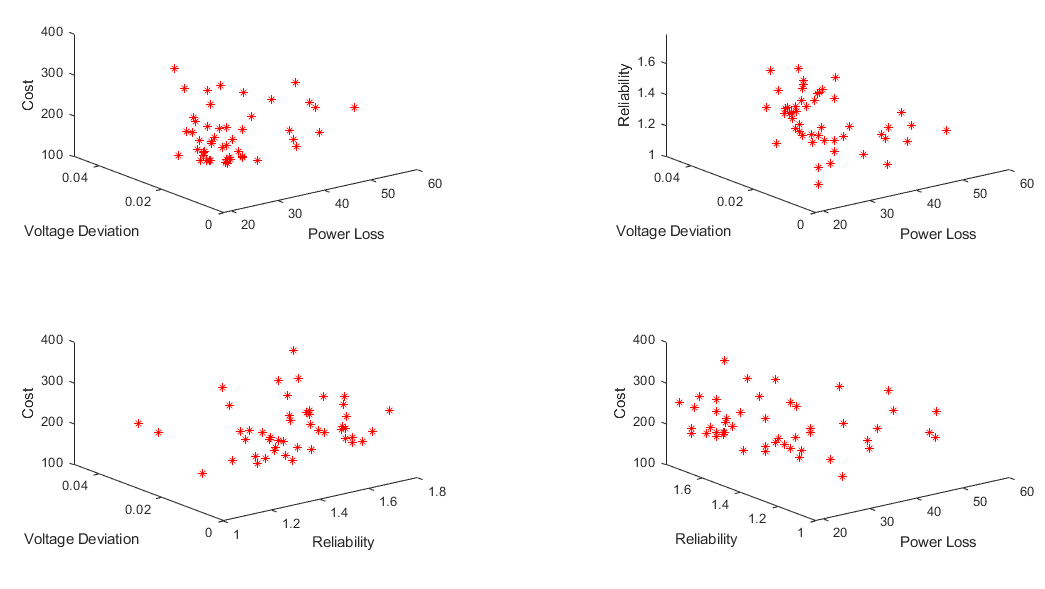}
			\caption{3-D plot of the Pareto optimal solutions for 3-objective function.}
			\label{fig.7}
		\end{center}
	\end{figure}

	To evaluate the algorithm presented for the DNR, we perform optimization with GA, PSO, and CSA  algorithms at 15 o'clock. The experiments were repeated 10 times and the results are presented in Table Table \ref{tab.6}. This shows that the proposed algorithm performs better in terms of CPU strength and time. The optimal results of the proposed Pareto model are given in Figure. \ref{fig.7}.
	\begin{table}
	\centering
		\caption{Reliability data of branch}
		\label{tab.5}
		\begin{tabular}{ *{4}{c} }
			\toprule
			\multirow{2}{1cm}{Component} & 	\multicolumn{3}{c}{Parameter}  \\
			\cline{2-4} \\
			& $ \textbf{failure rate} \lambda \left( f/year \right) $ & $ \textbf{MTTF} \left( h \right)  $ & $ \textbf{MTTR} \left( min \right)  $   \\
			\midrule
			$ \textbf{line} \left( 1km \right) $ 	 &  $ 0.128 $	   & $ 1/\lambda $	 &  $ 45 $\\
			\bottomrule
		\end{tabular}
	\end{table}
	\begin{table}
	\centering
		\caption{comparison of performance between GA, PSO, CSA and ICSA}
		\label{tab.6}
		\begin{tabular}{ *{5}{c} }
			\toprule
			\textbf{method} & \textbf{average} & \textbf{objective} &  \textbf{average} & \textbf{standard} \\
			& \textbf{CPU time} & \textbf{function}  &   & \textbf{deviation}  \\
			\toprule
			\multirow{4}{1cm}{\textbf{GA}} & \multirow{4}{1cm}{746.5978}  & \textbf{OF1} & 1.4579 & 0.4924  \\
			&& \textbf{OF2} & 28.9506   & 4.1625  \\
			&& \textbf{OF3} & 0.0162  &  0.0046  \\
			&& \textbf{OF4} & 212.3664  &  49.2960   \\
			\toprule
			\multirow{4}{1cm}{\textbf{PSO}} &\multirow{4}{1cm}{584.3626}  &  \textbf{OF1} & 1.4113 & 0.2815 \\
			&	& \textbf{OF2} & 28.6133 & 4.1830 \\
			&	& \textbf{OF3} & 0.0154  & 0.0043   \\
			&	& \textbf{OF4} & 200.8071  & 28.3773  \\	
			\toprule
			\multirow{4}{1cm}{\textbf{CSA}} & \multirow{4}{1cm}{489.9613} & \textbf{OF1} & 1.4112  & 0.2250 \\
			&	& \textbf{OF2} & 28.5176 & 3.4516 \\
			&	& \textbf{OF3} & 0.01508 & 0.0037  \\
			&	& \textbf{OF4} & 198.4572 & 24.7153  \\
			\toprule
			\multirow{4}{1cm}{\textbf{ICSA}} & \multirow{4}{1cm}{412.3084}  &  \textbf{OF1} & 1.4108 & 0.1413 \\
			&	& \textbf{OF2} & 28.3823 & 2.3784 \\
			&	& \textbf{OF3} & 0.0139  & 0.0025  \\
			&	& \textbf{OF4} & 193.1413 & 16.5915 \\
			
			\bottomrule
		\end{tabular}
	\end{table}
	\begin{figure}
		\begin{center}
			\includegraphics[height=6cm,width=10cm]{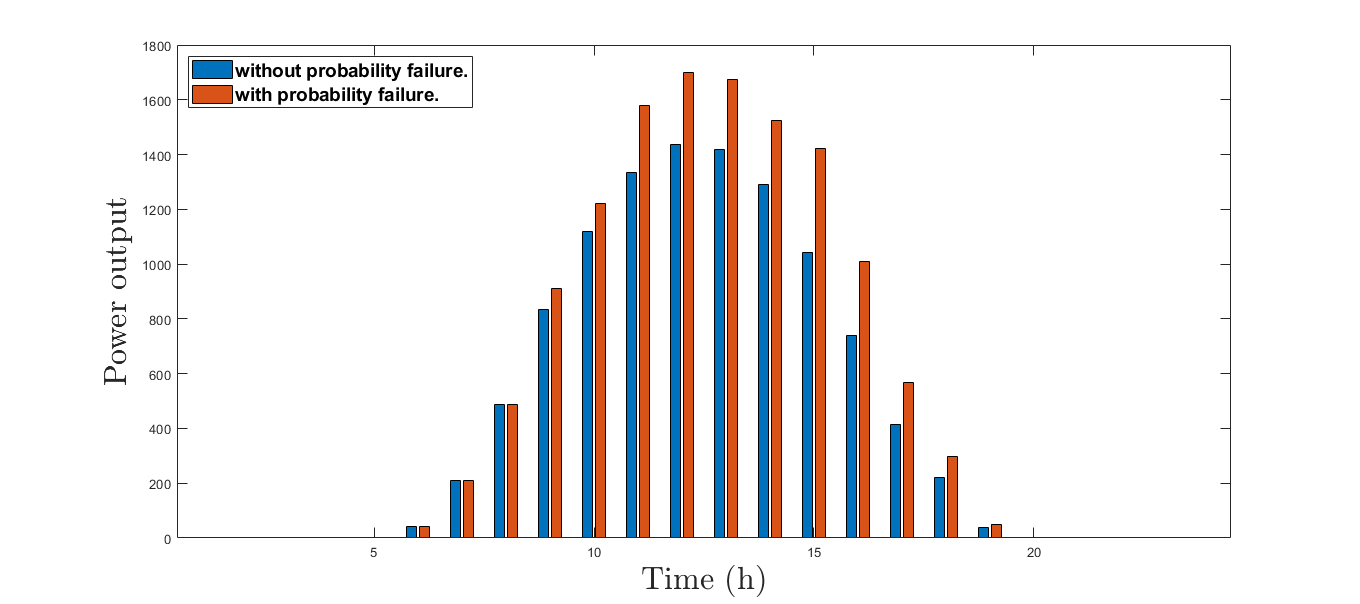}
			\caption{penetration of renewables with and without probability of failure.}
			\label{fig.9}
		\end{center}
	\end{figure}
	\begin{figure}
		\begin{center}
			\includegraphics[height=6cm,width=10cm]{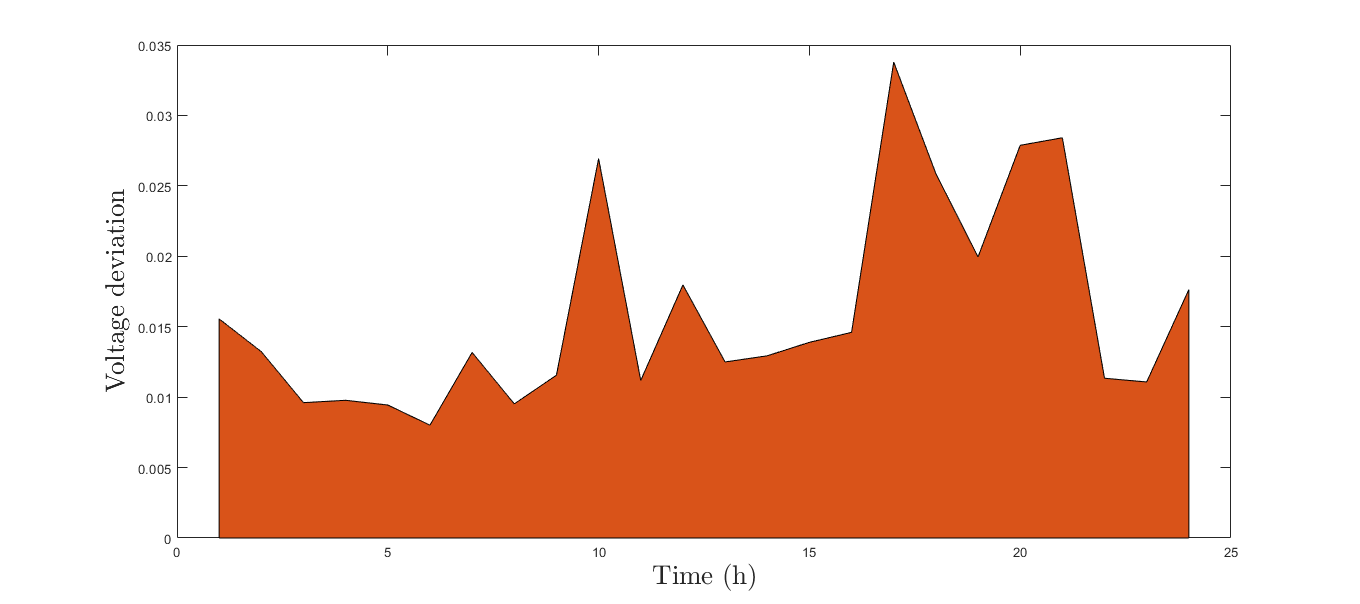}
			\caption{Maximum voltage deviation with probability of failure.}
			\label{fig.10}
		\end{center}
	\end{figure}
	\begin{table}
	\centering
		\caption{Average of objective functions in 24 hours, with and without probability of failure}
		\label{tab.11}
		\begin{tabular}{ *{3}{c} }
			\toprule
			\multirow{3}{1.5cm}{\textbf{objective function}}& \textbf{scenario 1}  & \textbf{scenario 2} \\
			& \textbf{with probability failure } & \textbf{without probability failure}  \\
			\cline{2-3}
			& \textbf{average} &  \textbf{average} \\
			\toprule
			\textbf{OF1} & 1.2758   & 1.4429 \\ 
			\textbf{OF2} & 27.4177  &  27.6733  \\
			\textbf{OF3} &  0.0161  &  0.0165  \\
			\textbf{OF4} & 246.3931 &  268.1462 \\
			\bottomrule
		\end{tabular}
	\end{table}
	\subsection{Checking the Possibility of Equipment Failure in SDNR} 
	1) Here are two scenarios for assessing dynamic reconfiguration, taking into account the uncertainty of load and generation. In the first scenario, the possibility of damage to renewable resources is not taken into account and it is assumed that they will not experience any change during the day. In the second scenario, random programming is done for renewable resources. The probability of source failures in optimization is considered. The number of failure sources in $ t_{0} $ in renewable farms PV1 to PV5 is 5, 0, 1, 2, and 3, respectively. As shown in Table \ref{tab.11}, considering the uncertainty effect of renewable resources, reconfiguration results in better performance in the objective functions. This means that in daily reconfiguration, DSM can improve its optimization goals by considering the possibility of repairing damaged equipment on the network.
	
	2) Figure. \ref{fig.9} shows that the penetration of renewables increases with the repair of damaged sections during the day. One of the major challenges of DNR in the presence of renewables is maintaining the consumer voltage within a limited range. As it can be seen in Figure. \ref{fig.10}, the proposed model has been able to control the maximum voltage deviation at distribution system, while accounting for the possibility of failure.
	
	\section{Conclusion} \label{sec.7}
	Given the high computation complexity of DNR, it is difficult to consider many issues such as equipment failure when performing DNR. In this paper, the possibility of failure of renewable resources is considered , leading to the so-called SDNR method. In this method, the number of scenarios is reduced using the probability distance method, then the best scenario is selected. An improved crow search algorithm (ICSA) has also been proposed to solve the DNR problem. The performance of this algorithm has been compared with the GA and PSO algorithms in terms of computation time. Due to the promising reduction in computation time, this algorithm is suitable for large-scale networks. There are multiple directions for future research on DNR. A short list of directions includes considering dispatch DG (at the same time as DNR), optimizing critical switches, and improving search methods to find the optimal solution.

\bibliographystyle{plain}
\bibliography{references}
\end{document}